\documentclass[aps,prb,twocolumn,floatfix,superscriptaddress,showpacs,%
showkeys,amssymb]{revtex4}
\usepackage{graphicx}
\usepackage{amssymb}

\begin{document}

\title{Spin-orbit coupling and intrinsic spin mixing in quantum dots}
\author{C. F. Destefani}
\affiliation{Department of Physics and Astronomy and Nanoscale and
Quantum Phenomena Institute, Ohio University, Athens, Ohio
45701-2979} \affiliation{Departamento de F\'{\i}sica, Universidade
Federal de S\~{a}o Carlos, 13565-905, S\~{a}o Carlos, S\~{a}o
Paulo, Brazil}
\author{Sergio E. Ulloa}
\affiliation{Department of Physics and Astronomy and Nanoscale and
Quantum Phenomena Institute, Ohio University, Athens, Ohio
45701-2979}
\author{G. E. Marques}
\affiliation{Departamento de F\'{\i}sica, Universidade Federal de
S\~{a}o Carlos, 13565-905, S\~{a}o Carlos, S\~{a}o Paulo, Brazil}
\date{\today }

\begin{abstract}
Spin-orbit coupling effects are studied in quantum dots in InSb, a
narrow-gap material.  Competition between different Rashba and
Dresselhaus terms is shown to produce wholesale changes in the
spectrum. The large (and negative) $g$-factor and the Rashba field
produce states where spin is no longer a good quantum number and
intrinsic flips occur at moderate magnetic fields.  For dots with
two electrons, a singlet-triplet mixing occurs in the ground
state, with observable signatures in intraband FIR absorption, and
possible importance in quantum computation.
\end{abstract}


\maketitle

The creation and manipulation of spin populations in
semiconductors has received a great deal of attention in recent
years.  Conceptual developments that have motivated these efforts
include prominently the Datta-Das proposal for a spin field-effect
transistor,\cite{1} based on the Rashba spin-orbit coupling of
electrons in a 2DEG,\cite{2} and the possibility of building
quantum computation devices using quantum dots ({\bf
QDs}).\cite{3} It is then important for full control of spin-flip
mechanisms in nanostructures that all spin-orbit ({\bf SO})
effects be understood.

There are two main SO contributions in zincblende materials like
$A_{3}B_{5}$: in addition to the structure inversion asymmetry
(\textbf{SIA}) caused by the 2D confinement (the Rashba effect),
there is also a SO term caused by the bulk inversion asymmetry
(\textbf{BIA}) in those structures (the Dresselhaus term).\cite{5}
Notice that additional lateral confinement defining a QD
introduces another SIA term with important consequences, as we
will see in detail. Although the relative importance of these two
effects depends on the materials and structure design (via
interfacial fields), only recently have authors begun to consider
the behavior of spins under the influence of all effects. For
example, a modification of the Datta-Das design was recently
suggested to allow for a diffusive version of the spin FET,
\cite{4} and that proposal relies on the additional influence of
the Dresselhaus SO coupling in the system.

Work in wide-gap materials (mainly GaAs), \cite{6} uses a unitary
transformation on the Hamiltonian of the system, \cite{7} after
which one gets an effective diagonal SO term which incorporates
the Rashba effect in a perturbative fashion.  That approach is
valid since the SO coupling is small in GaAs. However, the
approximation is not valid for all of the $A_{3}B_{5}$ structures,
as it is the case for InSb, for example, where both SIA and BIA
effects are anticipated to be much larger. \cite{Cardona} In this
material, one needs to deal with the full Hamiltonian.

There are just a few works discussing SO effects in narrow-gap
nanostructures. Among them, [\onlinecite{9}] uses $\mathbf{k \cdot
p}$ theory in InSb QDs in order to include SIA SO terms from both
the Rashba field {\em and} the lateral confinement which defines
the QD.  This last SIA term is considered in [\onlinecite{11}],
and since it is diagonal in the Fock-Darwin ({\bf FD}) basis no
level mixing is found nor expected. In contrast, level mixing
events are clearly identified in [\onlinecite{9}]. Experiments in
InSb QDs have explored the FIR response in lithographically
defined dots, \cite{10a} and PL features of self-assembled
dots.\cite{10b}

The goal of this work is to show how important different types of
SO couplings are in the spectra of parabolic QDs built in
narrow-gap materials such as InSb. We consider the Rashba-SIA
diagonal and SIA non-diagonal, as well as the Dresselhaus-BIA
terms in the Hamiltonian, and proceed with its full
diagonalization, in order to study features of the spectrum as
function of magnetic field, dot size, $g$-factor, and
electron-electron interaction.  We draw attention to the
appearance of strong level anticrossings (mixing) for moderate
magnetic fields in typical QDs, and how this phenomenon (and
`critical' field where it occurs) is modified by the BIA terms not
considered before.\cite{9}  As the level mixing involves states
with different spin, this induces strong {\em intrinsic} spin
flips in the system, {\em regardless} of the strength of the SO
coupling, providing an important channel for spin decoherence in
these systems.  Moreover, measurement of FIR absorption would
yield {\em direct} access to the coupling constants; i.e., the
dispersion of FIR absorption peaks and appearance of
additional/split-off features are a direct consequence of the
level mixing introduced by SO.

{\em Model}. Assuming a heterojunction or quantum well confinement
$V(z)$ such that only the lowest $z$-subband is occupied, the
Hamiltonian in the absence of SO interactions for a QD further
defined by a lateral parabolic confinement is given by $H_{0}=
\frac{\hbar ^{2}}{2m}\mathbf{k}^{2}+V(\rho )+\frac{1}{2}g\mu _{B}
\mathbf{B} \cdot \mathbf{\sigma }$, where
$\mathbf{k}=-i\mathbf{\nabla }+e\mathbf{A}/(\hbar c)$, and the
in-plane vector potential $\mathbf{A}=B\rho (-\sin \theta ,\cos
\theta ,0)/2$ describes a perpendicular magnetic field
$\mathbf{B}=B\mathbf{z}$; $m$ is the effective mass in the
conduction band,\cite{remark_non-parab} $g$ is the bulk
$g$-factor, $\mu _{B}$ is Bohr's magneton, $V(\rho )=\frac{1}{2}
m\omega _{0}^{2}\rho ^{2}$ is the lateral confinement with
frequency $\omega _{0}$, and in the Zeeman term $\sigma _{X,Y,Z}$
are the Pauli matrices. The analytical solution of $H_0$ yields
the FD spectrum with energies $E_{nl\sigma}=(2n+|l|+1)\hbar \Omega
+l\hbar \omega _{c}/2+g\mu _{B}B\sigma /2$, with effective
frequency $\Omega =\sqrt{\omega _{0}^{2}+\omega _{C}^{2}/4}$, and
cyclotron frequency $\omega _{C}=eB/(mc)$; states are given in
terms of associated Laguerre polynomials. \cite{FD} The
confinement, magnetic and effective lengths are, respectively,
$l_{0}=\sqrt{\hbar /(m\omega _{0})}$, $l_{B}=\sqrt{\hbar /(m\omega
_{C})}$ and $\lambda =\sqrt{\hbar /(m\Omega )}$.

The SIA SO term with coupling parameter $\alpha $ is
$H_{SIA}=\alpha \mathbf{\sigma }\cdot \left( \mathbf{\nabla
}V\times \mathbf{k}\right) $, where the total confinement
potential is $V(\mathbf{r})=V(\rho )+V(z)$.  One can then write
$H_{SIA}=H_{R}+H_{SIA}^{D}$, where the diagonal contribution
coming from the lateral confinement in cylindrical coordinates is
$H_{SIA}^{D}=\alpha \frac{\hbar \omega _{0}}{l_{0}^{2}}\sigma
_{Z}\left( L_{Z}+\frac{\lambda
^{2}}{l_{B}^{2}}\frac{x^{2}}{2}\right)$, with the adimensional
radial coordinate $x=\rho /\lambda $, and $L_{Z}=-i\partial
/\partial \theta $.

The Rashba term coming from the perpendicular confinement field
$dV/dz$ is
\begin{equation}
H_{R}=-\alpha \frac{dV}{\lambda dz}\left[ \sigma _{+}L_{-} A_- +
\sigma _{-}L_{+} A_+ \right]  \text{,}
\end{equation}%
where $L_{\pm }=\exp (\pm i\theta )$, $\sigma _{\pm }=(\sigma
_{X}\pm i\sigma _{Y})/2$, and operators $A_\pm = \mp \partial
/\partial x + L_{Z} /x + x\lambda ^2/(2l_{B}^{2})$.

In zincblende structures one should also consider the BIA SO bulk
Hamiltonian. \cite{5} After averaging in the $z$-direction, due to
quantization, one gets $H_{BIA}=\gamma \left[ \sigma
_{x}k_{x}k_{y}^{2}-\sigma _{y}k_{y}k_{x}^{2}\right] +\gamma
\left\langle k_{z}^{2}\right\rangle \left[ \sigma _{y}k_{y}-\sigma
_{x}k_{x} \right] +\gamma \sigma _{z}\left\langle
k_{z}\right\rangle \left( k_{x}^{2}-k_{y}^{2}\right) $, where
$\gamma$ is the coupling parameter, the resulting first (second)
term is cubic (linear) in the in-plane momentum, and the last term
is zero because $\left\langle k_{z}\right\rangle =0$; also,
$\left\langle k_{z}^{2}\right\rangle \simeq (\pi /z_{0})^{2}$,
where $z_{0}$ is the $z$-direction confinement length.  One may
write the BIA SO term as $H_{BIA}=H_{D}^{L}+H_{D}^{C}$, where the
linear Dresselhaus contribution is given by
\begin{equation}
H_{D}^{L}=i\frac{\gamma \left\langle k_{z}^{2}\right\rangle
}{\lambda } \left[ \sigma _{+}L_{+} A_+ - \sigma _{-}L_{-} A_-
\right] \text{,}
\end{equation}%
while the cubic contribution $H_{D}^{C}$ can be expressed in terms
of $\sigma_\mp L_\pm^3$ and $\sigma_\pm L_\pm$, and different
powers in $x$, $\partial/\partial x$, and $L_z$.
\cite{Destefani-thesis} Notice that under a finite magnetic field,
the matrix elements with $\sigma _{\pm}L_{\pm}$ in $H_{D}^{C}$ are
{\em not} hermitian, and one needs to symmetrize them; \cite{7} if
the field is zero, this problem does not occur. \cite{13A}

For the electron-electron interaction $H_{ee}$, an expansion in
Bessel functions for $|\mathbf{r}_{1} - \mathbf{r}_{2}|^{-1}$ is
employed. \cite{Destefani-thesis} The basis states are properly
antisymmetrized, describing the unperturbed spin eigenstates.

The general form of the various SO terms in the Hamiltonian
exhibit already interesting characteristics. For example, the
magnetic field plays a role via its linear dependence in
$H_{SIA}^D$, $H_R$, and $H_{D}^L$, or its $B$ to $B^3$ dependence
in $H_D^C$. \cite{Destefani-thesis} Most interestingly, this form
of the Hamiltonian yields selection rules explicitly, dictating
which levels will be influenced by the SO effects. For example, at
zero field the diagonal SIA term splits the levels according to
the total angular momentum $j$. The Rashba term induces a set of
anticrossings in the FD spectrum whenever $\Delta l=\pm 1 =
-\Delta \sigma$ at finite field (due to the $\sigma_\pm L_\mp$
terms; mostly negative $l$'s are affected since their magnetic
dispersions allow for crossings); the lowest anticrossing is
between $\{n,l,\sigma \}=\{0,0,-\}$ and $\{0,-1,+\}$.  The cubic
BIA terms (with $\sigma _{\mp }L_{\pm }^{3}$) induce a set of
anticrossings, which obey $\Delta l=\mp 3$ and $\Delta \sigma =\pm
1$; the first one at low $B$-field involves the states $\{0,1,-\}$
and $\{0,-2,+\}$. Terms with $\sigma _{\pm }L_{\pm }$ in $H_D^L$
and $H_D^C$ do not induce anticrossings, but rather split and
shift the spectrum due to matrix elements with $\Delta l=\pm
1=\Delta \sigma $.  Notice that the matrix elements between states
with different $n$'s are in general non-zero, so that the full
diagonalization involves mixings with various $n$-values.

{\em Results}. The sequence of FD states of $H_0$ starts at zero
$B$-field with $\{n,l,\sigma \} = \{0,0,\pm \}$, followed by the
degenerate set of $\{0,-1,\pm \}$ and $\{0,1,\pm \}$.\cite{FD}
Spin and orbital degeneracies are broken by $B$ and the states
with negative $l$ and positive $\sigma $ acquire lower energies
because of the {\em negative} $g$-factor. The lowest energy level
crossing is between states $\{0,0,-\}$ and $\{0,-1,+\}$, and the
field where it occurs in the FD spectrum is
\begin{equation}
B_{C}^0=\frac{\widetilde{m}}{\mu _{B}}\frac{\hbar
\omega_0}{\sqrt{\widetilde{m}|g|(\widetilde{m}|g|+2)}}\text{,}
\label{Bc}
\end{equation}
where $\widetilde{m}=m/m_0$. The moderate value of $B_C^0$ is a
direct consequence of the large $|g|$ in InSb.\cite{parameters}
For GaAs ($|g|=0.44$, $\tilde{m}=0.067$), for example, this level
crossing appears only at $B_C^{GaAs} \simeq 9.4$T for a much
smaller confinement, $\hbar \omega_0 = 2$meV, and in the region
where Landau levels are well defined. Weaker confinement (smaller
$\omega_0$) shifts this crossing to lower fields.  Notice that for
$g <0$, $H_R$ mixes these states (and $H_D^L$ shifts the crossing
to higher fields).  For $g>0$ it is $H_D^L$ that would produce
relatively stronger level anticrossings (and $H_R$ would only
shift the spectrum weakly), and it would then be absent in
non-zincblende materials like silicon.

The energy spectrum for InSb QDs with typical characteristics,
\cite{parameters} and for the full Hamiltonian is presented in
Fig.\ 1A vs.\@ $B$ field.  The spectrum is obtained by direct
diagonalization using a FD basis with $n \leq 4$ (or ten energy
`shells'), i.e. 110 basis states.  We have studied the progressive
changes to the FD levels when including different SO terms in $H$.

\begin{figure}[tbp]
\includegraphics*[width=8.9cm]{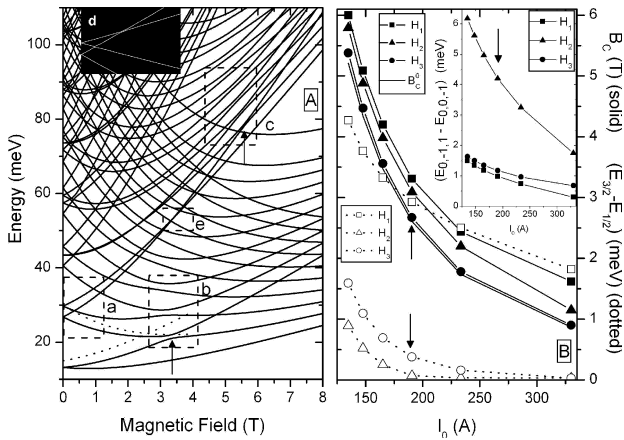}
\caption{A. Full Hamiltonian $H$ spectrum vs.\ $B$ field for InSb
QD as in [\onlinecite{parameters}]. Highlights in dashed boxes:
{\sf a} shows zero-field splitting in second shell (affected
mostly by BIA SO terms), and crossing at about $0.3T$; compare
with inset {\sf d} with only SIA terms and {\em two} crossings at
$0.02$ and $0.06T$, and much smaller zero-field splitting. Second
crossing for this shell in full $H$ is at $3.4T$ ({\sf e} box).
{\sf b} and {\sf c} indicate anticrossings (AC) induced by Rashba
term with $\Delta l=-\Delta\sigma=\pm1$; first AC (arrow) in {\sf
b} involves states $\{0,0,-\}$ and $\{0,-1,+\}$ ($\{0,1,-\}$ and
$\{1,0,+\}$ in {\sf c}). Dotted lines indicate FD levels crossing
at $2.6T$.
B. Lateral size dependence. Dotted lines: SO zero-field splitting
in {\sf a} box on left panel. Solid lines: $B_C$ field of first AC
in {\sf b} box on left panel; inset shows splitting at that AC.
Arrows at 190\AA\, show QD size for spectrum in A. $H_{1}$ curves
(squares) use [\onlinecite{parameters}]; $H_{2}$ (triangles) and
$H_{3}$ (circles) use same parameters but four times stronger
Rashba field ($H_{2}$) or twice as large $z_0$ ($H_{3}$). Both
cases increase relative strength of SIA terms. Solid line with no
symbol shows $B_{C}^{0}$ in (\ref{Bc}).} \label{fig1}
\end{figure}

The diagonal $H_{SIA}^{D}$ term shifts energies but does not
change appreciably the position of the first crossing (\ref{Bc}),
shown by the dotted lines in Fig.\ 1A at $\simeq 2.6$T.\@ The
energy shifts induce two new crossings at low fields (inset {\sf
d}), since the SO orders states according to their total angular
momentum $j=l+s$; the highest (lowest) state at zero field has
$j=3/2$ $(1/2)$ in the second shell. At about 0.2T one recovers
the `normal' sequence of states: $\{0,-1,+\}$, $\{0,-1,-\}$,
$\{0,1,+\}$, $ \{0,1,-\}$. This competition between SO and
magnetic field is similar to the Zeeman and Paschen-Back regimes
in atoms. \cite{Paschen-Back-ref} We should note that this level
ordering is observed in [\onlinecite{11}].

The non-diagonal Rashba contribution $H_{R}$ introduces strong
state mixing for {\em any} value of the $\alpha$ parameter
whenever FD levels with $\Delta l=-\Delta \sigma = \pm 1$ cross.
This mixing converts the crossings at $B_C^0$ to clear
anticrossings. Higher levels which satisfy these selection rules
also anticross at nearly the same field.  The field-width and
energy-amplitude (or level splitting) of the mixing is dictated by
the value of $\alpha $, while the value of $B_C$ where the
anticrossing occurs is nearly unaffected by $\alpha$.

The cubic Dresselhaus contribution $ H_{D}^{C}$ induces
anticrossings (via $\sigma _{\mp }L_{\pm }^{3}$) and zero-field
splittings ($\sigma _{\pm }L_{\pm }$) in the FD spectrum.  The
splittings are much smaller than those induced by the Rashba term
and practically unnoticeable in the spectrum, reflecting the
smallness of the $ E_{D}^{C}$ for these parameters.
\cite{parameters}  For the linear BIA contribution, however, the
$\sigma _{\pm }L_{\pm }$ terms in $H_{D}^{L}$ have a much bigger
impact on the zero-field splittings, which can in principle be
`tuned' by changing the effective $z$-size, $z_0$. $H_{D}^{L}$
alone induces such a strong mixing at low fields that one cannot
identify the two Zeeman and Paschen-Back regimes.

Notice in the full spectrum of $H$ (Fig.\ 1A) that the first group
of anticrossings (for $n=0$ levels) induced by $H_{R}$ is shifted
to higher field due mostly to $H_{D}^{L}$, so that $B_C^0
\rightarrow B_{C}\simeq 3.3$T (box {\sf b} and lower arrow). The
set of anticrossings at $\simeq 5.5$T is also due to $H_R$ and
arises from the $n=1$ level manifold (box {\sf c} and upper
arrow).  At low field, only a single crossing in the second shell
at $\simeq 0.3$T is present and dominated by $H_{D}^{L}$ (box {\sf
a}; compare with inset {\sf d} and notice second crossing in box
{\sf e}). The sequence of the first excited levels at zero field
is $j=3/2$ $(1/2)$ for higher (lower) energy, while at higher
energies both SIA and BIA terms cooperate to produce anticrossings
(not visible at the resolution in Fig.\ 1A).

\begin{figure}[tbp]
\includegraphics*[width=8.9cm]{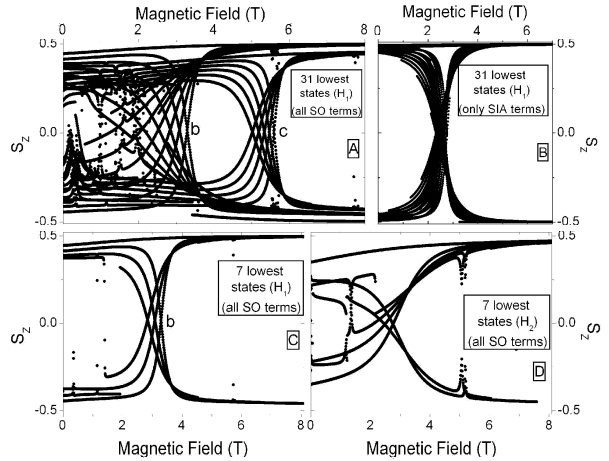}
\caption{A: Spin $z$-component vs.\ $B$ for lowest 31 states; {\sf
b} and {\sf c} labels refer to boxes in Fig.\ 1A. Higher energy
anticrossings in each set are shifted to lower fields. If only SIA
terms are considered (panel B), all spin mixing occurs at field $B
\simeq 2.6$T.
C: $S_z$ for lowest 7 states; full spin mixing at anticrossing. D:
dot with Rashba field 4 times stronger ($H_{1}$ and $H_{2}$
defined as in Fig.\ 1). Increasing SIA SO produces stronger
mixing.} \label{fig2}
\end{figure}

Figure 2 illustrates the importance of the level anticrossings on
the spin, as the expectation value of $S_z$ for each state is
plotted vs.\ $B$.  Figs.\ 2A and B include all states with $E
\lesssim 80$meV (for full SO and only SIA terms, respectively),
while figs. 2C and D focus only on the lowest seven levels.
Although a large majority of states have $S_z$ close to $\pm 1/2$,
as one expects for pure states, there are significant deviations.
The various SO terms mix levels close to accidental degeneracy
points in the FD spectrum and produce the large deviations seen in
the figure. 2C shows how $H_R$ produces an {\em intrinsic} (i.e.,
no phonon-assisted) total collapse of the spin number for the low
energy states in the QD.\@ Although the ground state is nearly
pure ($S_z \simeq 1/2$, and more so at higher $B$), the first few
excited states totally mix at $B_C \simeq 3.3$T.  2D shows how a
stronger Rashba field ($dV/dz = -2 \times 10^{-3}$eV/\AA) greatly
widens the mixing region and lowers $B_C \simeq 2.8$T.

One can further appreciate the intricate balance of SO terms under
a magnetic field. We analyze how various quantities are affected
by changes in the the lateral and vertical sizes, $l_{0}$ and $
z_{0}$, or the Rashba field $dV/dz$, as shown on Fig.\ 1B. The
zero-field splitting (dotted lines) is dominated by the linear BIA
contribution for any value of $l_{0}$ here. Increasing $z_{0}$
strongly reduces the splittings because the Dresselhaus
contribution weakens; the reduction is even more drastic if one
increases $dV/dz$, which makes the $H_{R}$ contribution bigger and
can then cancel or suppress better the splitting produced by
$H_{SIA}^{D}$. Some authors have considered the possibility of
tuning such SO terms to produce total cancellation of the
zero-field splitting, although considering only $H_{R}$ and
$H_{D}^{L}$.\cite{4} However, one also has to take into account
$H_{SIA}^{D}$ and $H_D^C$ contributions, which may be important
(the zero-field cancellation occurs at values of $z_0$ or Rashba
field about ten percent smaller than with only the former terms).
One should notice, in any event, that this change in parameters
only eliminates the zero-field splitting but not the anticrossing
at finite field, and measurement of both quantities on the same
sample could yield information on the {\em relative} strength of
the $\alpha$ and $\gamma$ parameters.

The anticrossing field $B_{C}$ (solid lines/symbols) decreases
with QD size, roughly according to (\ref{Bc}), $B_C^0 \simeq
\omega_0 \simeq 1/\sqrt{l_0}$.  A finite $\alpha $ slightly
increases $B_{C}$, but the BIA contribution considerably upshifts
it, as mentioned above. Increasing $z_{0}$ or $dV/dz$ decreases
$B_{C}$. At $l_{0}=320$\AA\, ($\hbar \omega_{0}=5$meV), $
B_{C}=1.6$T, while it shifts to 1.15T if $dV/dz$ is four times
larger or to 0.85T if $z_{0}$ is doubled, both cases decreasing
the BIA contribution.  These values are comparable to those in
[\onlinecite{9}] without including BIA terms (adjusting for
differences in system parameter values). Anticrossings at such low
fields may be interesting for applications due to easier access.

The energy splitting at $B_{C}$ (inset in Fig.\ 1B), has main
contribution from the Rashba term for any dot size considered, but
the BIA reduces the splitting substantially. If $z_{0}$ is changed
from $40$ to 80\AA\, the splitting is enhanced slightly, but
larger $z_0$ produces no significant changes. However, the
splitting is drastically enhanced if one increases the Rashba
field.  Here, the splitting goes from 1 to 4.2meV if the interface
field is increased fourfold.

\begin{figure}[tbp]
\includegraphics*[width=8.9cm]{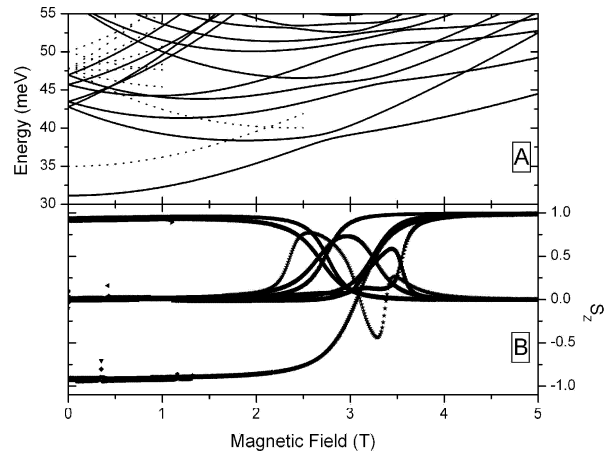}
\caption{A: Two particle spectrum vs.\ $B$ field for full
Hamiltonian $H+H_{ee}$ (basis included 190 states, only lowest
levels shown). With no SO (dotted lines) and at $B=0$, ground
state is a singlet ($\{L,S\} =\{0,0\}$) at 35meV while first
(second) excited state is a triplet ($\{\pm1,\pm1\}$ and
$\{\pm1,0\}$) at 48meV (singlet ($\{\pm1,0\}$) at 50meV).  SO acts
against electron-electron interaction, as levels are shifted back
to energies close to non-interacting case. Lowest anticrossing at
$\simeq 2.7$T is between singlet ground state $\{0,0\}$ and lowest
excited triplet state $\{-1,1\}$; SO introduces coupling between
the singlet and triplet states with direct consequences for QD
ground state.
B: Spin $S_z$ for the nine lowest states of the two particle QD.
Strong mixing induced by SO interaction appears in all states
$\simeq 3$T.} \label{fig3}
\end{figure}

Figure 3 illustrates the corresponding level structure for two
electrons in the QD (full Hamiltonian $H+H_{ee}$; dashed lines
show $H_0+H_{ee}$, the non-SO case). The repulsive interaction
shifts the ground state upwards by $\simeq 5$meV, and the exchange
shifts the triplet down by 2meV. Most interestingly, the SO
interaction introduces a strong mixing of the singlet and triplet
transition at $B\simeq 2.7$T. The fact that the mixing occurs at
relatively low field makes that a possibly useful transition for
the implementation of quantum computing devices. Moreover, the
splitting will also be apparent in the FIR response of QDs,
allowing the determination of the various SO coupling strengths.

We have shown that inclusion of all SO terms is essential in order
to obtain a complete picture of the level structure in narrow-gap
QDs.  The combination of strong SO couplings and large (and
negative) $g$ factor introduces strong intrinsic mixing of the low
excitations for the single-particle spectrum.  Consequently, the
two-particle spectrum exhibits strong singlet-triplet coupling at
moderate fields, with significant experimental consequences.
Observation of FIR mode magnetic dispersion would allow the direct
determination of coupling constants.

\vspace*{1ex} We acknowledge support from FAPESP-Brazil, US DOE
grant no.\ DE-FG02-91ER45334, and the CMSS Program at OU.


\begin{thebibliography}{10}
\bibitem{remark_non-parab}  Non-parabolicity effects are important
in InSb but as they do not introduce spin mixing, we have ignored
them in this work.  Their effect is to shift energy levels
somewhat, which will in turn shift the critical field $B_C$.
\end{thebibliography}
\end{document}